\documentclass[article]{aa} 

\newcommand{\grad}{$^{\circ}$}

\usepackage{blindtext}
\usepackage{graphicx}
\usepackage{float}
\usepackage{multirow}
\usepackage{txfonts}
\usepackage{subfig}
\usepackage{hyperref}
\usepackage{soul}
\usepackage{color}
\usepackage{amsmath}
\usepackage{makecell}
\hypersetup{
    colorlinks=true,
    linkcolor=red,
    filecolor=magenta,      
    urlcolor=cyan
    }
\begin{document} 
   
    \title{Revisiting the Einstein Cross with X-shooter spectra: New estimates of accretion disk parameters and broad emission line profiles from microlensing}
    
\author{R. Jerez-Nicurcar\inst{1}
        \and  A. Esteban-Guti\'{e}rrez\inst{1, *}
        \and  V. Motta\inst{1, *}
        \and  E. Mediavilla\inst{2,3}
        }
    \institute{Instituto de Física y Astronomía (IFA), Facultad de Ciencias, Universidad de Valparaíso, 
             Av. Gran Bretaña 1111, Playa Ancha, Valparaíso, Chile.  \\
     * Corresponding authors, \email{ana.esteban@uv.cl}, \email{veronica.motta@uv.cl}
            \and{Instituto de Astrofísica de Canarias, Vía Láctea s/n, La Laguna 38200, Tenerife, Spain}
            \and{Departamento de Astrofísica, Universidad de la Laguna, La Laguna 38200, Tenerife, Spain}
            }    

\abstract{We present a spectroscopic study of the lensed quasar Q2237+0305, the Einstein Cross, assessing the best wavelength coverage to date with the X-shooter instrument (from $3000$ to $20000 \ \AA$), and thus improving the available spectroscopic data for this system. Using as a reference the characteristic emission lines for each quasar macro-image, we estimate the chromatic microlensing effect between pairs of images and calculate the extinction curve, confirming no significant differences between spectra $C$ and $A$ caused by extinction in the lens galaxy. In addition, we model the accretion disk characterized by a temperature profile with logarithmic slope $1/p$ and size $r_{s}$, assuming a thin-disk model with a power-law function of the form $r_{s}$ $\propto$ $\lambdaup^{p}$. To do this, we calculated a set of three chromatic microlensing measurements compiling previous estimates from the literature to test the wavelength dependence. This temperature profile was also compared to the canonical thin-disk model with $p=4/3$. Finally, we estimated the likelihood between the disk model and the observed data collected from multiple epochs through mock microlensing magnification maps and determined the probability density functions of each parameter from a Bayesian analysis. As a result, our estimated values for the accretion disk size and its exponential index are $r_{s}=4.9^{+2.3}_{-2.6}$ $\sqrt{M/0.3M_{\odot}}$ (lt-day) and $p=0.77^{+0.47}_{-0.28}$, respectively, reinforcing a departure from the standard accretion disk model ($ T \propto r_s^{-3/4}$).}


\maketitle

\section{Introduction}

Accessing the internal structure of active galactic nuclei (AGNs) remains a major observational challenge, as these systems are extremely distant and characterized by very small angular scales. Over the past several decades, reverberation mapping (RM) has been the primary technique used to probe the innermost regions of AGNs, providing direct constraints on the size and structure of the broad line region (BLR) through the time delay between continuum and line emission \citep{peterson1993,bentz2009,Fian2022,Fian2023}. Despite its success, RM requires long-term, high-cadence monitoring campaigns, which limits its applicability to a relatively small number of nearby or well-monitored sources.

An alternative and complementary approach is offered by gravitational microlensing in lensed quasars. When a quasar (QSO) is multiply imaged by a foreground galaxy, stars in the lens can induce microlensing magnification events that selectively affect emission regions of different physical sizes. This phenomenon introduces extrinsic flux variability between the QSO macro-images that is sensitive to angular scales on the order of microarcseconds \citep{1998Wambsganss}. As a result, microlensing provides a powerful tool to probe the inner structure of AGNs using single-epoch spectroscopy, as has been demonstrated by numerous previous studies (e.g., \citealt{motta2012,rojas2020,fian2021,Melo2023,2024Hutsemekers}).

Thus, when microlensing is present in one or more lensed images, the QSO spectrum can exhibit wavelength-dependent magnification of the continuum emission. This chromatic behavior allows for one to characterize the accretion disk size and temperature profile by comparing differential spectra between images, even from single-epoch observations. Such analyses have yielded estimates of the parameters of the accretion disk and are used to test different models and investigate whether there is a universal description of QSO accretion disks \citep{eigenbrod2008II,2011Mosquerab,mediavilla2011,2011Blackburne,sluse2012,motta2017}. 

Moreover, microlensing can also affect the emission line profiles of QSOs. Since the BLR is composed of regions with different velocities, microlensing can induce flux anomalies in specific parts of a broad emission line (BEL), particularly in its wings \citep{lewis1998,abajas2002,sluse2011,fian2021}. In contrast, larger emitting regions are less susceptible to microlensing. Consequently, the line cores, which originate from more extended regions of the BLR, are commonly used as a reference baseline unaffected by microlensing, while the continuum emission is typically more strongly magnified.

Among known lensed quasars, the system Q2237+0305 (also known as the Einstein Cross) stands out as an exceptional laboratory for microlensing studies. This system consists of a nearby lensing galaxy at redshift $z_d = 0.039$ and a background quasar at $z_s = 1.695$ \citep{huchra19852237}. The four macro-images are formed within the bulge of the lens galaxy, resulting in a high optical depth to microlensing and an unusually frequent occurrence of microlensing magnification events compared to other lensed QSOs.

Q2237+0305 has therefore been the subject of extensive long-term photometric and spectroscopic monitoring campaigns, with extrinsic flux variations predominantly attributed to microlensing. Early spectroscopic observations revealed flux anomalies between images, notably in the Mg\,{\sc ii} emission line \citep{filippenko1989}, later confirmed by \cite{irwin1989}. One of the first systematic photometric monitoring efforts was carried out by the OGLE collaboration \citep{wozniak2000}, which reported significant and rapid brightness variations in the quasar images. Since then, numerous studies have investigated the microlensing variability and structural properties of this system using both photometric and spectroscopic data \citep{wyithe2000, Yonehara2001,anguita2008,poindexter2010,sluse2011,mediavilla2015,Goicoechea2020,berdina2021}.

Spectro-photometric analyses have further demonstrated the diagnostic power of microlensing in Q2237+0305. \cite{eigenbrod2008} detected strong chromatic microlensing in the continuum emission of image $A$ using \texttt{VLT/FORS1} data, finding that the inner, hotter regions of the accretion disk are more strongly magnified than the cooler, outer regions. This wavelength-dependent magnification is consistent with a temperature profile similar to that proposed by \cite{kochanek2004}. Subsequent studies applied detailed chromatic microlensing models to constrain the accretion disk size and temperature profile \citep{sluse2012}, while more recent works have focused on characterizing both the accretion disk and the BLR using different modeling approaches and continued monitoring \citep{2021hutse,savic2024,sorgenfrei2024}. However, despite the extensive literature devoted to Q2237+0305, a fully consistent picture of its inner structure remains elusive. Existing studies often rely on heterogeneous datasets, limited wavelength coverage, or focus on individual microlensing events. In particular, constraints derived from chromatic microlensing are frequently restricted to the optical regime, while the impact of differential extinction between images and its interplay with microlensing effects are not always treated in a unified manner. Therefore, a coherent analysis combining broad wavelength coverage, homogeneous spectroscopy, and a systematic comparison with previous microlensing measurements is still lacking.

In this work, we aim to address these limitations through a comprehensive and homogeneous analysis of the Einstein Cross, based on archival multi-wavelength \texttt{VLT/X-shooter} spectroscopy combined with multiband photometric data from the literature. Thus, our goals are: (1) to provide an improved characterization of the BLR structure by extending the analysis from the UV to the NIR, (2) to estimate the differential extinction curve between the quasar images, and (3) to compile and reanalyze published chromatic microlensing measurements together with our new constraints extending into the near-infrared, in order to refine previous accretion disk size and temperature profile estimates. By combining these elements within a unified framework, this work provides a more robust view of the inner structure of Q2237+0305 and its relevance as a benchmark system for microlensing studies of AGNs.

The structure of the paper is as follows. Section 2 presents the observational data used for the analysis of Q2237+0305 along with a description of the spectral reduction for the \texttt{X-shooter} instrument data. Section 3 describes the analysis performed on the emission lines found in the \texttt{X-shooter} spectra and the calculation of the differential magnitude measurements. Section 4 provides a detailed methodology for the computation of microlensing maps and the statistical approach used to extract the properties of the accretion disk, followed by a description of the results in Section 5. Finally, Section 6 summarizes the conclusions of this work.

\section{Observational data}
\subsection{X-shooter spectral data}

\begin{figure}
    \centering
    \includegraphics[width=0.48\textwidth]{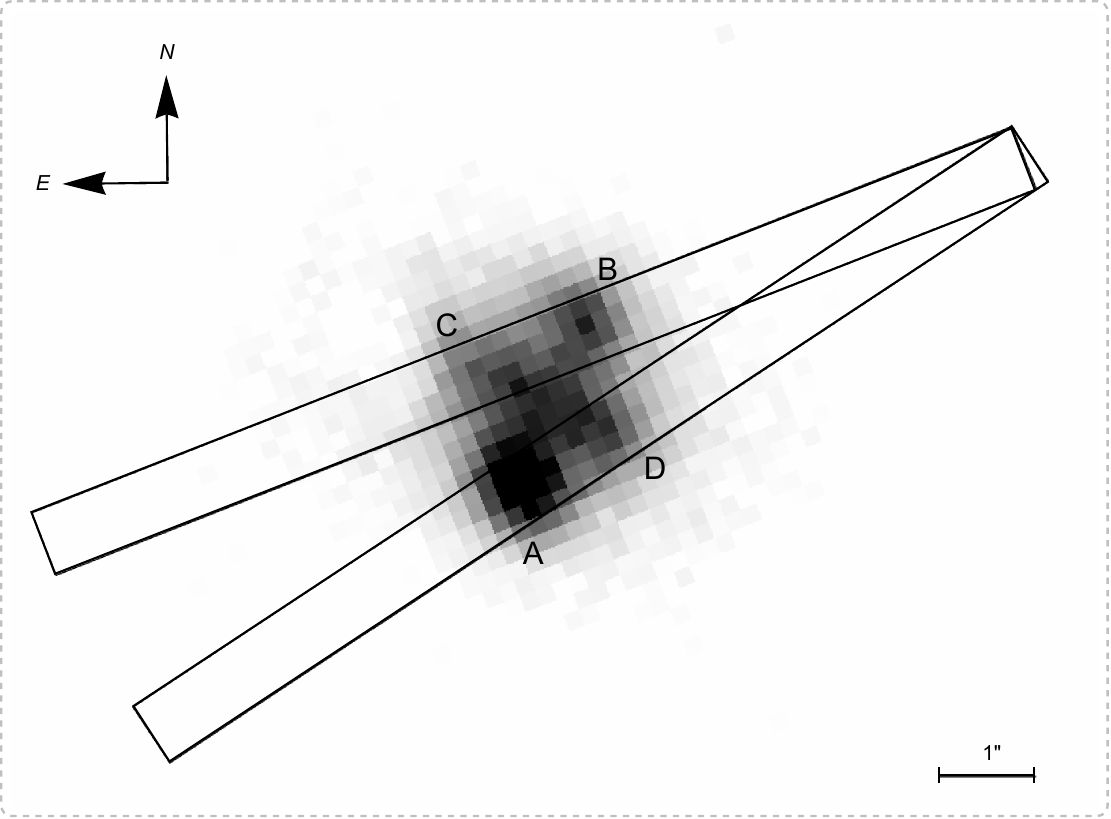}
    \caption{\texttt{X-shooter} acquisition image of Q2237+0305 in the r band. The slits used for the VIS arm are shown (11"$\times$0.7") over the $AD$ and $BC$ quasar images. The quasar images are labeled as $A$, $B$, $C$, and $D$ according to the literature, with $A$ being the brightest image in this epoch.}
    \label{fig:slit_q2237} 
\end{figure}

We collected spectra of the four lensed images of Q2237+0305 from the ESO archive\footnote{\href{http://archive.eso.org/cms.html}{http://archive.eso.org/cms.html}} of the \texttt{X-shooter} instrument \citep{xshooter} mounted at the Very Large Telescope (VLT)-UT2, taken in June 2018, with proposal ID 101.B.033A (P.I: Lira, P.). \texttt{X-shooter} is a medium-resolution Echelle Spectrograph that covers wavelengths from 3000 to 24800 $\AA$ and is separated in three arms (UVB, VIS, and NIR). The slit was centered on the brightest image, with a position angle such as it covered two out of the four images at a time. This slit position allows us to obtain the spectra in two groups: $AD$ and $BC$ lensed quasar images (see Figure \ref{fig:slit_q2237}). The size of the UVB, VIS, and NIR slits were $11"\times0.8"$ (R=6700),  $11"\times0.7"$ (R=11400), and $11"\times0.6"$ (R=8100), respectively, in order to obtain a similar wavelength resolution of $0.02,0.02$, and $0.06 \ \AA/pix$ for each arm. Observing information for this system is provided in Table \ref{tab:Q2237_log}.

\begin{table*}[h!]
\caption{Observing information of Q2237+0305 with the \texttt{X-shooter} instrument.}
\label{tab:Q2237_log}
\centering
\small
\renewcommand{\arraystretch}{1.5}{
\begin{tabular}{lllllllll} \hline\hline
Images & Obs. Date$^{a}$        & P.A$^{b}$.      &  Arm       & Exp. time$^{c}$       & Slit  & Seeing
& FWHM & Airmass$^{d}$\\
       & (yy-mm-dd)  & (\grad)      &              & (s)     &   ('')& ('')   & ('') &  \\ \hline\hline
       &             &       &   \small UVB        & 8 x 323   & 11 x 0.8 & 0.86 & 0.82 &  1.53   \\ 
 $AD$    &  2018-06-20 &  56.5 &   \small VIS        & 8 x 322   & 11 x 0.7 & 0.80 & 0.81 & 1.53       \\ 
       &             &       &   \small NIR        & 8 x 79    & 11 x 0.6 &  0.74 &  0.67 &1.53\\ \cline{4-9}
       &             &       &   \small UVB        & 8 x 323   & 11 x 0.8 &  1.26 & 1.18 &1.62\\
$BC$     & 2018-06-20  & 68.5  &   \small VIS        & 8 x 322   & 11 x 0.7 & 1.23 & 1.30  &1.62\\
       &             &       &   \small NIR        & 8 x 79    & 11 x 0.6 & 1.34 & 1.24 &1.62\\\hline\hline
\end{tabular}\\
}
\begin{flushleft}
\footnotesize{\textbf{Notes:}}\\
\footnotesize{$^a$ Observation date.\\
$^b$ Slit position angle from north to east.\\
$^c$ Exposure time of each observing frame. \\
$^e$ \textit{Airmass} at the beginning of the observation night.}\\
\end{flushleft}
\end{table*}

\begin{figure*}[t!]
    \centering
    \hspace{-1em}
    \includegraphics[width=0.85\textwidth]{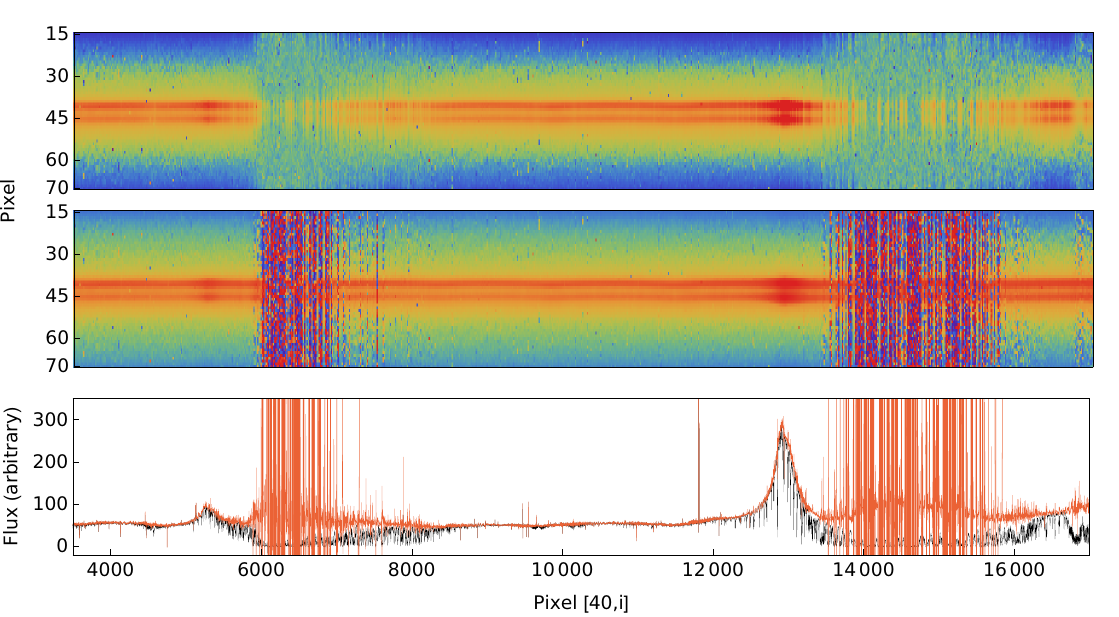}
    \caption{Spectra telluric corrections. Top and middle panels: Comparison of the spectra before and after telluric correction using \texttt{molecfit} in the NIR \texttt{X-shooter} arm, respectively. Bottom panel: 1D spectrum around component $A$ with (orange) and without (black) telluric correction.} 
    \label{fig:spec_2d_telluric}
\end{figure*} 

\subsubsection{Data reduction}

\indent The data processing and calibration were carried out with the \texttt{X-shooter} pipeline version 3.5.0 \citep{xshooter} in the \texttt{EsoReflex} environment\footnote{ESO Recipe Flexible Execution Workbench} \citep{reflex}. Data were bias and flat-field corrected, sky-subtracted, and wavelength-calibrated. Among the different configurations given by the pipeline, we used the \textit{nodding} option to reduce the data and the \textit{stare} one to estimate the sky dispersion, respectively. In the \textit{nodding} mode, we can automatically subtract the sky emission by taking advantage of the A and B nodding positions (within the slit), which are separated by 3.5" from the slit center. 
We manually localized the spectra and using the pipeline we set the slit aperture centered at the halfway of the two spectra (either \textit{AD} or \textit{BC}), taking widths of 2.5" for the UVB and VIS arms and 2.0" for the NIR. 
The \textit{stare} mode was used to confirm that the available windows for the sky subtraction were sufficient to perform the correction. 
We did not find any differences in the spectra between the two methods.

The telluric correction in the VIS and NIR arms of the 2D spectra was applied row by row using the \texttt{molecfit} \citep{smette2015molecfit} package. 
To calculate the \texttt{molecfit} best-fit function, the frames were mean-combined into a single spectrum for each arm, allowing an increment in the signal-to-noise ratio (S/N). A comparison of the NIR arm before and after this telluric correction is shown in Figure \ref{fig:spec_2d_telluric}.  
Finally, no flux calibration was performed since we were only interested the flux ratio between the images to measure the microlensing signal.

\subsubsection{Spectra extraction}

\indent 

Considering the spectra were taken under different seeing conditions (see Table \ref{tab:Q2237_log}), we use a Moffat\footnote{$M(x) = A_i \cdot (1 + (x-\bar{x}_i)^2/\sigma^2 )^{-\beta}$, where $\bar{x}_i$ is the position and $A_i$ the amplitude and subindex $i$ denotes each individual Moffat function fitting for every QSO image, while $\sigma$ and $\beta$ are the dispersion and broadening parameters, respectively. The FWHM of a Moffat function is estimated as $2\sigma\cdot \sqrt{1/2^{1/\beta} -1}$.} function to correct from cross-contamination between image pairs in the slit and perform a fit for each QSO image. Regarding the contamination of the lens galaxy, given the configuration of the system (Figure \ref{fig:slit_q2237}), we assume that it is the same for all components. 

We employed a stacking technique centered on the  C{\footnotesize{III}}], Mg{\footnotesize{II}}], and H$\alpha$ emission line wavelengths, for the UVB, VIS, and NIR arms, respectively, to obtain the initial estimates for the centroids, amplitudes, and dispersion ($\sigma$) parameters as of the Moffat function. 
Finally, a 1D spectrum for each QSO image was extracted by fixing the centroids and the $\sigma$ parameter (assumed to be the same for both components), with $A_{i=1,2}$ (two amplitudes in each QSO image) and $\beta$ as free parameters.

\begin{figure}
   \centering  \includegraphics[width=0.5\textwidth]{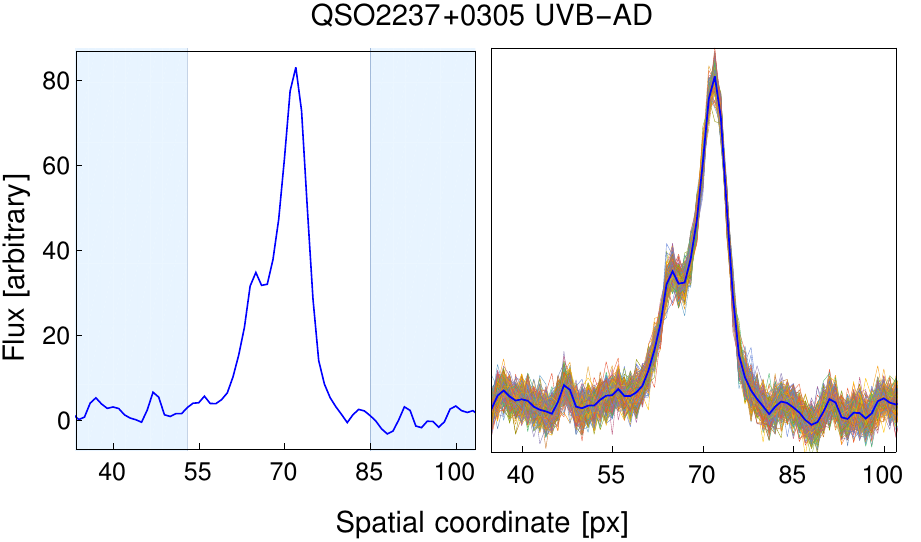}
    \caption{UVB $A-D$ one-pixel sample spectra. Left: Sky selected areas are shown in light blue corresponding to the sample spectra (solid blue line). Right:
Set of 300 simulated spectra using a Monte Carlo approach.}
    \label{fig:simulated_spectra}
\end{figure}

The fitting uncertainties were estimated with a Monte Carlo approach \citep{montecarlo}.  We chose a lower S/N spectra sample and performed a resampling of our selection by introducing white noise. This noise is estimated as the sky dispersion in two spatial regions on either side of a low-S/N region of the spectra, as we can see in Figure \ref{fig:simulated_spectra} (left panel). The white noise was incorporated into the sample spectra as a random value using a normal distribution centered at zero with the value of the sky dispersion as the standard deviation. To calculate the statistically significant number of simulations, we used the formula $n(\log n)^2$, where $n$ is the degree of freedom. As we considered an interval of $\sim70$ pix for each line, the minimum simulation number required was $\sim 300$ (Figure \ref{fig:simulated_spectra}, right panel). We applied our fitting method and estimated the error in the parameters at a $90\%$ confidence level.

\begin{figure*}[t!]
    \centering
    \hspace{-1em}
     \includegraphics[width=0.94\textwidth]{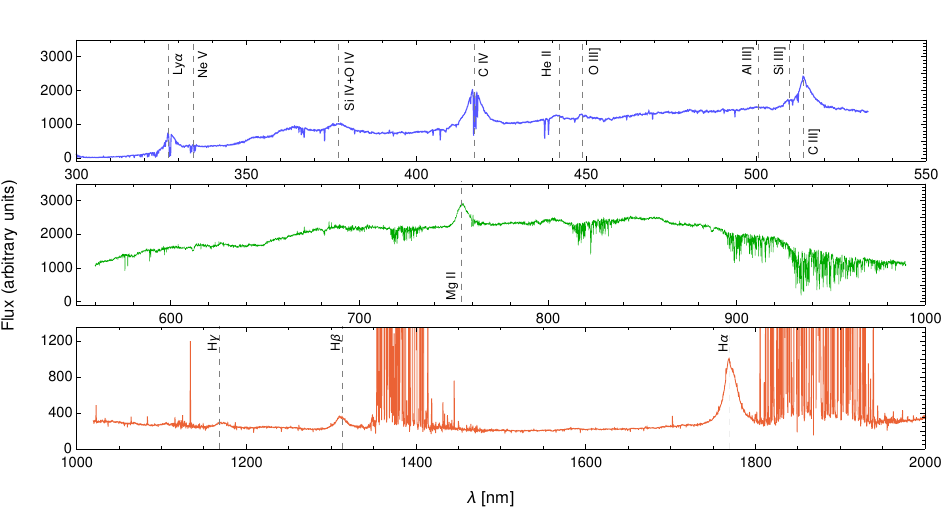}

     \vspace{-1.25em}\includegraphics[width=0.94\textwidth]{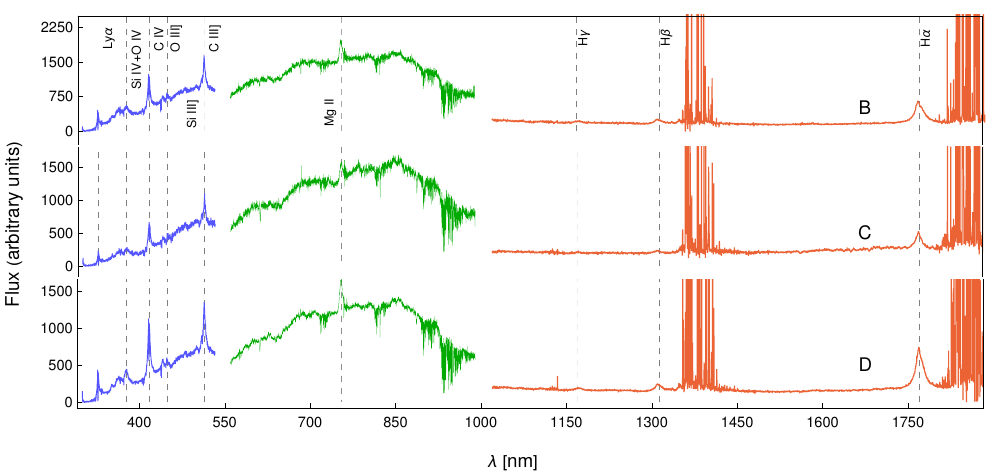}
    \caption{\texttt{X-shooter} extracted spectra for Q2237+0305. Top panel: Spectra for the $A$  image in the three different instrument arms UVB (blue), VIS (green), and NIR (red). The spectra is in the quasar rest frame ($z_{s}=1.695$) and the observed emission lines are marked with dashed gray lines. Bottom panel: Extracted spectra for the $B$, $C$, and $D$ images with the joint instrument arms.}
    \label{fig:ABCD_spectrum} 
\end{figure*}

The uncontaminated spectra for each QSO component were obtained by integrating the individual Moffat function at each wavelength, providing the flux ($F$) with its corresponding uncertainty ($\delta F$).
Finally, we obtained the magnitude uncertainties for the flux extraction using $|\delta m_e| \approx |\delta F / F |$. The final uncontaminated 1D spectra extracted for the four Q2237+0305 images can be seen in Figure \ref{fig:ABCD_spectrum}. For a better visualization of the emission lines, the first panel shows the individual zoom-in of each spectral arm (UVB: blue; VIS: green; NIR: orange) for image $A$, while the bottom panel corresponds to a merge of the three used \texttt{X-shooter} spectral arms for images $B$, $C$, and $D$, taking into account that the gaps represent the overlapping of the spectral arms.

\subsection{Multi-epoch photometric data}

For extending the sample of differential microlensing measurements between the Q2237+0305 image pairs, we collected a data set which covers photometric observations obtained from CASTLES\footnote{The CfA-Arizona Space Telescope LEns Survey} \citep{CASTLES,falco1999} with \textit{The Hubble Space Telescope} (HST), from the UV to the NIR bands added to narrow-band photometry data from the Nordic Optical Telescope (NOT) located in the Roque de los Muchachos Observatory, La Palma, for seven narrow filters plus a wide I-Bessel
\cite[see Table 2 from][]{mosquera2009}
for a wavelength range of 3510-8130 $\AA$. These additional data were used in the estimation of the accretion disk parameters, further detailed in Section 4.

\section{Analysis of Q2237+0305 spectra}
\subsection{Continuum fitting and flux estimation}\label{flux_estimation} 
The procedure of measuring the microlensing signal consists of obtaining the difference between the flux ratios of the continuum (prone to microlensing) and the core of the emission lines, which is believed not to be affected by microlensing \citep{mediavilla2009,mediavilla2011,guerras2013}, for image pairs ($B-A$, $C-A$, and $D-A$). The continuum was obtained by performing a linear fit in two regions on either side of each emission line. We selected the core of emission lines to avoid contamination by the wings of the BEL, which could also be affected by microlensing
(\citealt{abajas2002,eigenbrod2008,sluse2012,guerras2013,motta2017,fian2018,rojas2020, melo2021}). Thus, the magnitude differences measured in the core of the emission lines ($\Delta m^{core}$) can be used as a reference baseline \citep{rojas2014,motta2017,melo2021}. 

\begin{figure*}
\centering
 \includegraphics[width=0.85\textwidth]{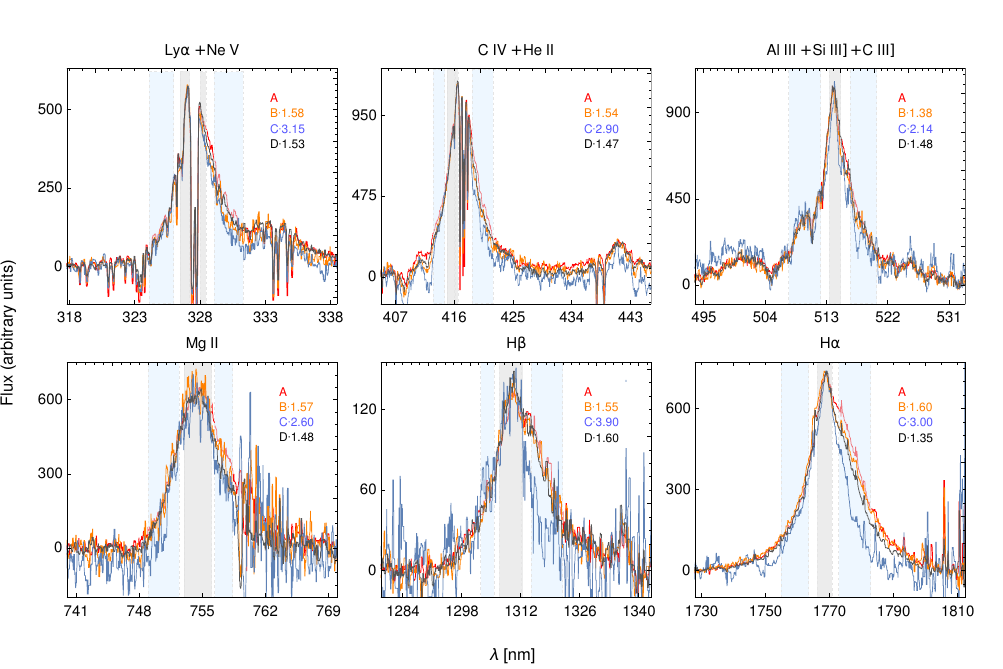}
  \caption{Matched profiles of the emission lines found with the \texttt{X-shooter} spectra for the Q2237+0305 system. The different colors represent the profiles associated with each quasar image and the values in the legend indicate the normalization factor used to match the core of the  faintest emission line to the brightest one ($A$). The gray and light blue bands correspond to the integration window of the line core and the BEL, respectively. For a better visualization of the profiles, a Gaussian filter was applied.}
  \label{fig:lines_matched_Q2237}
\end{figure*}

As a particular case, when examining the individual lines, we noticed that the Mg{\footnotesize{II}} has been affected by telluric absorption close to its emission line core and that the H$\beta$ and H$\alpha$ lines have only been affected in the reddest part of the BEL, as can be seen in Figure \ref{fig:lines_matched_Q2237}. For these lines, we carefully selected the wavelength ranges when subtracting the continuum, avoiding the telluric regions and considering a wider continuum region in the bluest part than in the reddest part. In general terms, for all the emission lines we selected a continuum region following the  pseudo-continuum suggested by \citep{mejia-restropo2016}, being the NIR part of the spectrum more difficult to fit due to the low S/N, specially for image $C$.

To estimate the continuum flux below the emission line, we applied $F_{cont}=(a/2)(\lambdaup_{f}^2-\lambdaup_{i}^2)+b(\lambdaup_{f}-\lambdaup_{i})$ (where $a$ and $b$ represent the slope and the ordinate axis intersection, respectively) and performed the integral of the linear fit between $\lambdaup_{i}$ and $\lambdaup_{f}$.
The uncertainties in ($a$,$b$) were estimated as the standard deviation of the parameters of the linear function model and these were used to estimate the flux uncertainties of the continuum fit as $\sigma_{F} =(\Delta a/2)(\lambdaup_{f}^2-\lambdaup_{i}^2)+\Delta b(\lambdaup_{f}-\lambdaup_{i})$, where $ \delta m_f \simeq \sigma_F/F_{cont}$. Finally, the total magnitude uncertainty was obtained as $\delta m=\sqrt{(\delta m_e)^2+(\delta m_f)^2}$. Although not included explicitly in Figure \ref{fig:lines_matched_Q2237} to avoid saturation in the plot, typical errors in the flux for the emission lines were found to range between $\sim4\%$ and  $\sim 25\%$, the H$\beta$ lines and H$\alpha$, MgII for image $C$ being the most affected ones.

Once the continuum emission is subtracted, we obtained the magnitude of the core of the emission line by integrating in a narrow window (spanning 5-70 $\AA$ depending on each case) centered on the peak of the line.
In the case of $\mathrm{Ly\alpha}$ and C{\footnotesize{IV}} lines, both NELs were affected by absorption lines; thus, it was necessary to select two alternative regions as close to the peak as possible. 
The wavelength ranges of these chosen windows (continuum and line core) are shown in Table \ref{table:dif_Q2237}. As the estimated core flux relies on the continuum fitting, its uncertainty is the one calculated for the continuum.
Finally, we estimated the flux ratios between the image pairs for the continuum and the NEL and calculated the magnitude differences (Table \ref{table:dif_Q2237}). 
The microlensing signal between two images, at a given wavelength, was then obtained as the differential magnitude measurements between the continuum and the core for each image pair as $\Delta m=(m{_i}-m{_A})_{cont}-(m{_i}-m{_A})_{core} \equiv \Delta m_{iA}^{cont} - \Delta m_{iA}^{core}$, $i=B,~C,~D$, and $A$ being the reference image.

\subsection{Comparison of the broad emission line profiles}

After fitting the spectra, we could analyze the profile of the emission lines to search for microlensing signal in the wings of the BEL (see Figures \ref{fig:slit_q2237} and \ref{fig:spec_2d_telluric}). This was done by fitting and  subtracting the continuum below the emission line and matching their cores for all four images. 
 If any difference in the broad line profile is found, an integration window was used in both red and blue wings to obtain the flux and magnitude differences, so that we could get an estimation of the size of the BLR, as carried out by other previous works \citep{sluse2011,guerras2013,motta2017,rojas2020,fian2018,fian2021}. 

\subsection{Dust extinction law}

Considering that the light of the quasar images is crossing different dust and gas sections within the lens galaxy, the received total flux could be affected by differential extinction. 
As dust extinction is the only phenomenon that can affect the magnitude difference in the core of the emission lines, these can be used to estimate the dust extinction law in the lens galaxy \citep{falco1999,2002motta,2011mediavilla_ext,rojas2020} by applying the following equation:

\begin{equation}
    m_{i}(\lambdaup) - m_{A}(\lambdaup) = -2.5log \left(\frac{\mu_{i}}{\mu_{A}}\right)+(E_{i}-E_{A})R_{V}\left(\frac{\lambdaup}{1+z_{L}}\right)
,\end{equation}

\noindent where $\mu_{i}/\mu_{A}$ is the magnification ratio between the pair of images $ i=B,C,D$ and $A$, $ (E_{i}-E_{A})$ is the differential extinction coefficient and $R_{V}(\lambdaup/(1+z_{L}))$ is the extinction curve in the lens galaxy rest frame. For this calculation, we applied a Milky Way-type extinction curve with an extinction coefficient of $R_{V}=3.1$ \citep{cardelli1989}.

\section{Modeling of microlensing magnification maps: Accretion disk estimations}

To estimate the accretion disk temperature profile and the characteristic size of its emitting region, we mimicked the microlensing conditions by computing microlensing magnification maps for each quasar image using the inverse polygon mapping (IPM) algorithm from \cite{mediavilla2006,mediavilla2011}, which takes as inputs the parameters of the macro-model given by \cite{kochanek2004}, presented in Table \ref{tab:modeling_Q2237}, along with random positions of a number of microlenses of a given mass in the lens plane (note that the magnification probability is related to the mean mass of the microlenses, as previously determined by \citealt{schechter2004}). These simulated maps were then compared to our microlensing measurements (see \citealt{jimenez-vicente2014} and references therein) from the spectral analysis along with those from the multi-epoch photometry described in Section 2.2. To perform this, we applied a Bayesian statistical approach by modeling the accretion disk with a Gaussian intensity profile through the formula $I(R) \propto \exp(-R^2/2r_s^2)$, where $r_{s}$ is the size of the accretion disk and can be related to the half-light radius as $R_{1/2}=1.18r_{s}$ \citep{mortonson2005}. This parameter depends on the wavelength range and can be parameterized by a power law, $r_{s}\propto \lambdaup^{p}$, where $p$ is the power-law index and the value $p=4/3$ corresponds to the standard thin-disk model (S$\&$S; \citealt{Shakura1973}).

\begin{table}[H]
\centering
\caption{SIE lens model parameters for Q2237+0305.}
\begin{tabular}{ccccc}\hline
Q2237+0305 &  A  & B & C & D \\\hline \hline
$\kappa$ & 0.394  & 0.375 & 0.743 & 0.635 \\
$\gamma$ & 0.395 & 0.390 & 0.733 & 0.623 \\\hline
  \end{tabular}
\label{tab:modeling_Q2237}
\end{table}

Using the IPM, we created a magnification map for the four quasar images by randomly distributing stars in the maps of mean mass $M=0.3 M_\odot$ (representative mean mass of the Salpeter distribution) and taking into account that all linear scales in the disk size can be re-scaled with the mass as: $r_{s}\propto \sqrt{M/0.3M_\odot}$). The maps have a 2000$\times$2000 px${^2}$ resolution, with a pixel size of 0.58 lt-day, covering a 30$\times$30 Einstein radii area. According to the selected mass, the Einstein radius of the microlenses is $R_{E} = 38.9$ lt-day. For our measurements, we assumed a $\Lambda$ cold dark matter cosmological model with values $\Omega_{\Lambda}=0.7$, $\Omega_{M}=0.3$, and H$_{0}$= 70 kms$^{-1}$ Mpc$^{-1}$. On the other hand, since the quasar images are formed in the bulge of a spiral galaxy, we consider that all the matter is in the form of compact objects; that is, the smooth matter density fraction is zero in our calculations ($\kappa = \kappa_{star}$). The values for the convergence $\kappa$ and shear $\gamma$ parameters were collected from \citep{kochanek2004} and are shown in Table 2. 

\begin{figure}[h!]
    \centering
    \includegraphics[width=0.4\textwidth]{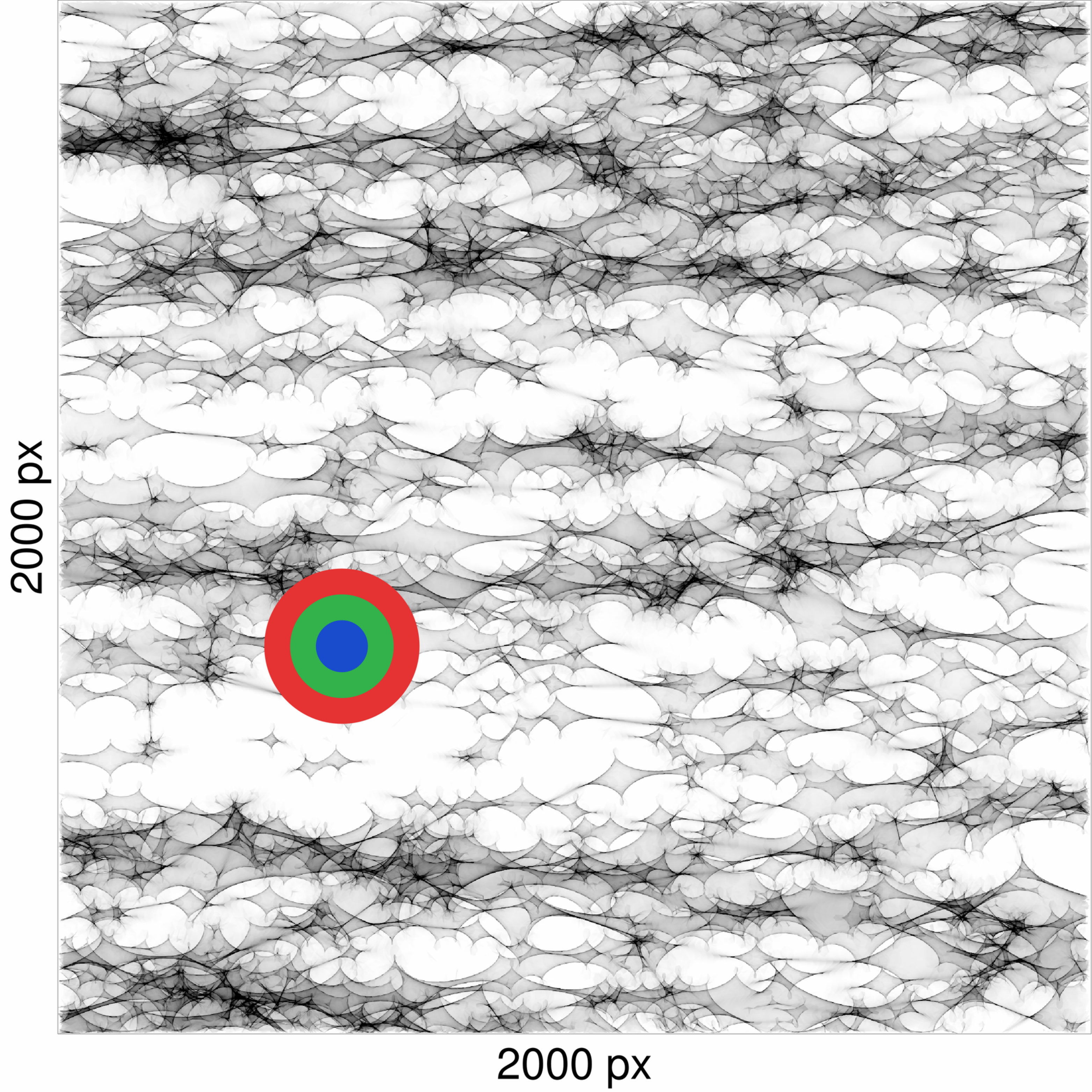}
    \caption{Microlensing magnification map for image A of Q2237+0305. Circles represent the source with different sizes, which depend on the three different wavelengths chosen (from red to blue: $\lambdaup=$754, 516, and 327 nm). We expect that a smaller source (blue) would be more affected by microlensing than a larger source (red).}
    \label{fig:map_mag}
\end{figure}

Once we modeled the microlensing magnification map for each quasar image (see an example of image $A$ in Figure \ref{fig:map_mag}), we obtained an estimation of the simulated magnitudes differences by convolving the maps with the sizes of our parametrized accretion disk,
$r_{s}(\lambdaup_{k})=r_{s}(\lambdaup_{k}/\lambdaup_{ref})^{p}$, at three different wavelengths,
$\lambdaup_{k}= \{327.0,~516.0,~754.0 \ nm\}$. The characteristic size was allowed to scale with an arbitrary power-law index, $p$, to test alternative models of the temperature profile departing from the S$\&$S. For comparison, we selected a reference value of $\lambdaup_{ref} = 1026~\AA$ (rest-frame) as the bluest wavelength for which microlensing magnifications were applied in the calculations by \cite{jimenez-vicente2014}.
To test the likelihood, we covered a linear grid of $(p,\ln{r_s})$ space parameters, where $p=0.15\cdot i$ ($i$ from 0 to 16), and $ \ln r_{s}=0.15\cdot j$ ($j$ from 0 to 40). 
The final likelihood of these three magnifications at each wavelength ($\Delta m_{k}^{obs}=\Delta m^{cont}-\Delta m^{core}$) for each pair of values ($p, \ln{r_s}$) can be calculated as

\begin{equation}
P_{l}\left(\Delta m_{k}^{\mathrm{obs}} \mid p_{i}, \ln r_{s j}\right) \propto \int \Delta m_{1} \int \Delta m_{2} \int \Delta m_{3} N_{a b} e^{-\frac{1}{2} \chi^{2}}
,\end{equation}

\noindent where $\Delta m_{1}$, $\Delta m_{2}$, and $\Delta m_{3}$ are the simulated magnitude differences measured at the three wavelengths, $N_{ab}$ is the number of iterations for each pair of pixels between the two maps, $a$ and $b$ (typically we chose $N_{ab}=10^8$), and $\chi^{2}$ is defined by

\begin{equation}
\chi^{2}=\sum_{k=1}^{3} \frac{\left(\Delta m_{k}^{\mathrm{obs}}-\Delta m_{k}\right)^{2}}{\sigma_{k}^{2}}
,\end{equation}

\noindent with $\sigma_{k}^{2}$ being the uncertainties measured at the three wavelengths chosen from our chromatic microlensing estimations (see Table \ref{table:chromatic_microlensing}).

Depending on the strength of the chromatic microlensing effect, the single-epoch method conducted on specific systems can be either overestimating or underestimating the accretion disk size \citep{guerras2017,2020guerras}. This can be solved by multiplying the likelihoods of multiple epochs of a single system or, alternatively, of multiple systems with similar sizes of the accretion disks as discussed in \cite{jimenez-vicente2012}. Here we used the former case, collecting $M$ different measurements for $p$ and $r_{s}$, as

\begin{equation}
P\left(\Delta m_{k, l}^{\mathrm{obs}} \mid p_{i}, \ln r_{s j} \right) \propto \prod_{l=1}^{M} P_{l}\left(\Delta m_{k}^{\mathrm{obs}} \mid p_{i}, \ln r_{sj} \right),
\end{equation}

\noindent where the indices $k$ and $l$ correspond to each lambda or chromatic measurement and the collection of different data to compare the simulations as seen in Table \ref{table:chromatic_microlensing}, respectively. 

\section{Results and discussion}

We obtained the uncontaminated spectra for the four quasar images in the UVB, VIS, and NIR \texttt{X-shooter} arms (Figure \ref{fig:ABCD_spectrum}). The spectra show prominent emission lines covering a wide wavelength range. Within them, we detected: Ly$\alpha$  $\lambdaup$121.5, Si{\footnotesize{IV}}]+O{\footnotesize{IV}}] $\lambdaup$139.9, C{\footnotesize{IV}} $\lambdaup$154.9, C{\footnotesize{III}}] $\lambdaup$190.8, Mg{\footnotesize{II}}] $\lambdaup$279.9, H$\beta$ $\lambdaup$486.3 y H$\alpha$ $\lambdaup$656.5 nm, and other ones with less flux, N{\footnotesize{V}} $\lambdaup124.1$, He{\footnotesize{II}} $\lambda164.0$, [O{\footnotesize{III}}] $\lambdaup165.5$, Al{\footnotesize{III}} $\lambdaup185.7$, Si{\footnotesize{III}}] $\lambdaup189.6$, O{\footnotesize{II}} $\lambdaup372.7$, H$\gammaup$ $\lambdaup434.2$ y [O{\footnotesize{III}}] $\lambdaup496.0$ nm, also marked in Figure \ref{fig:ABCD_spectrum} with dashed gray lines. 

\subsection{Continuum microlensing}\label{microlensing}
\begin{figure*}[t!]
    \centering
    \includegraphics[width=0.322\textwidth]
    {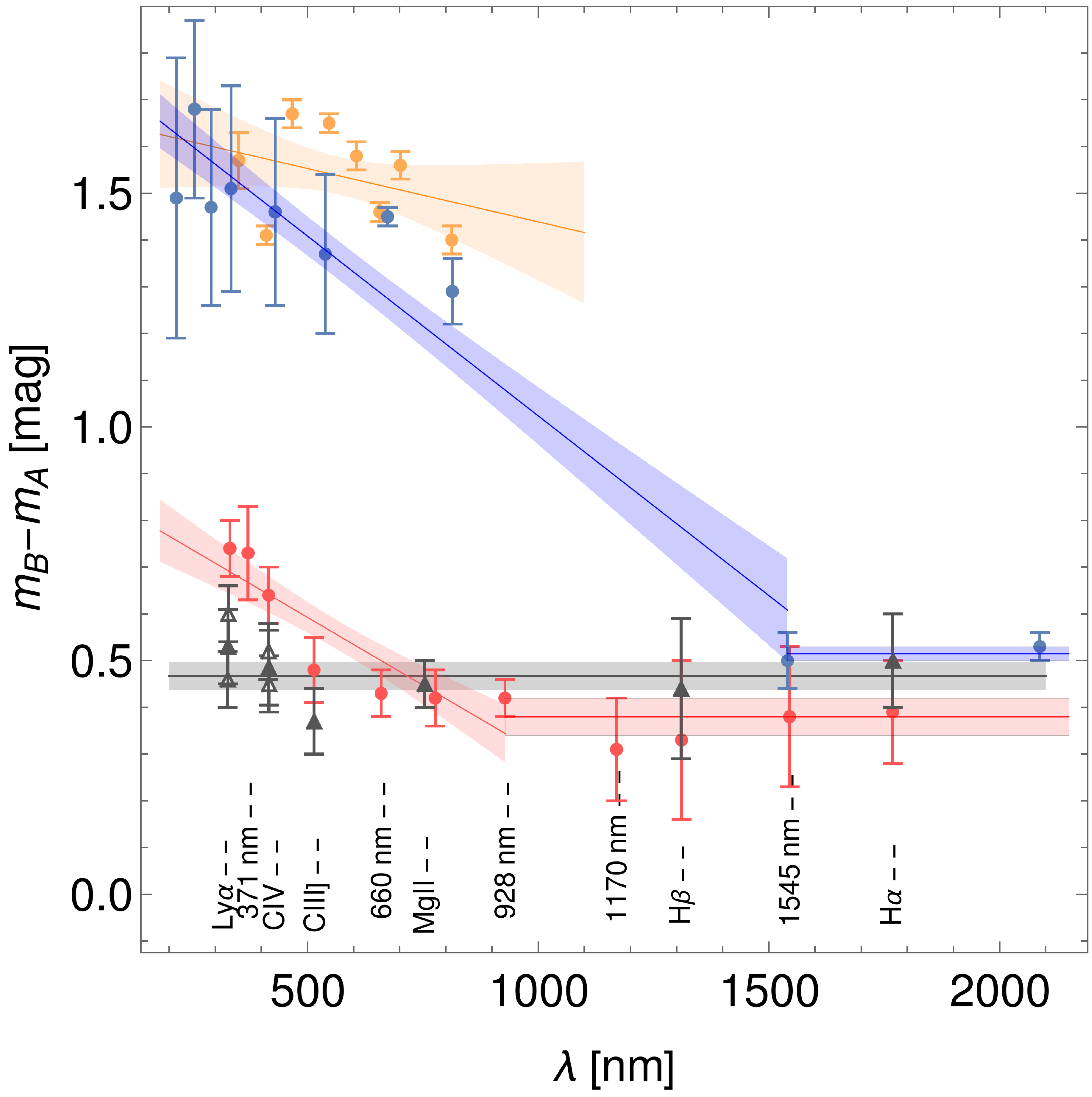}
    \includegraphics[width=0.33\textwidth]
    {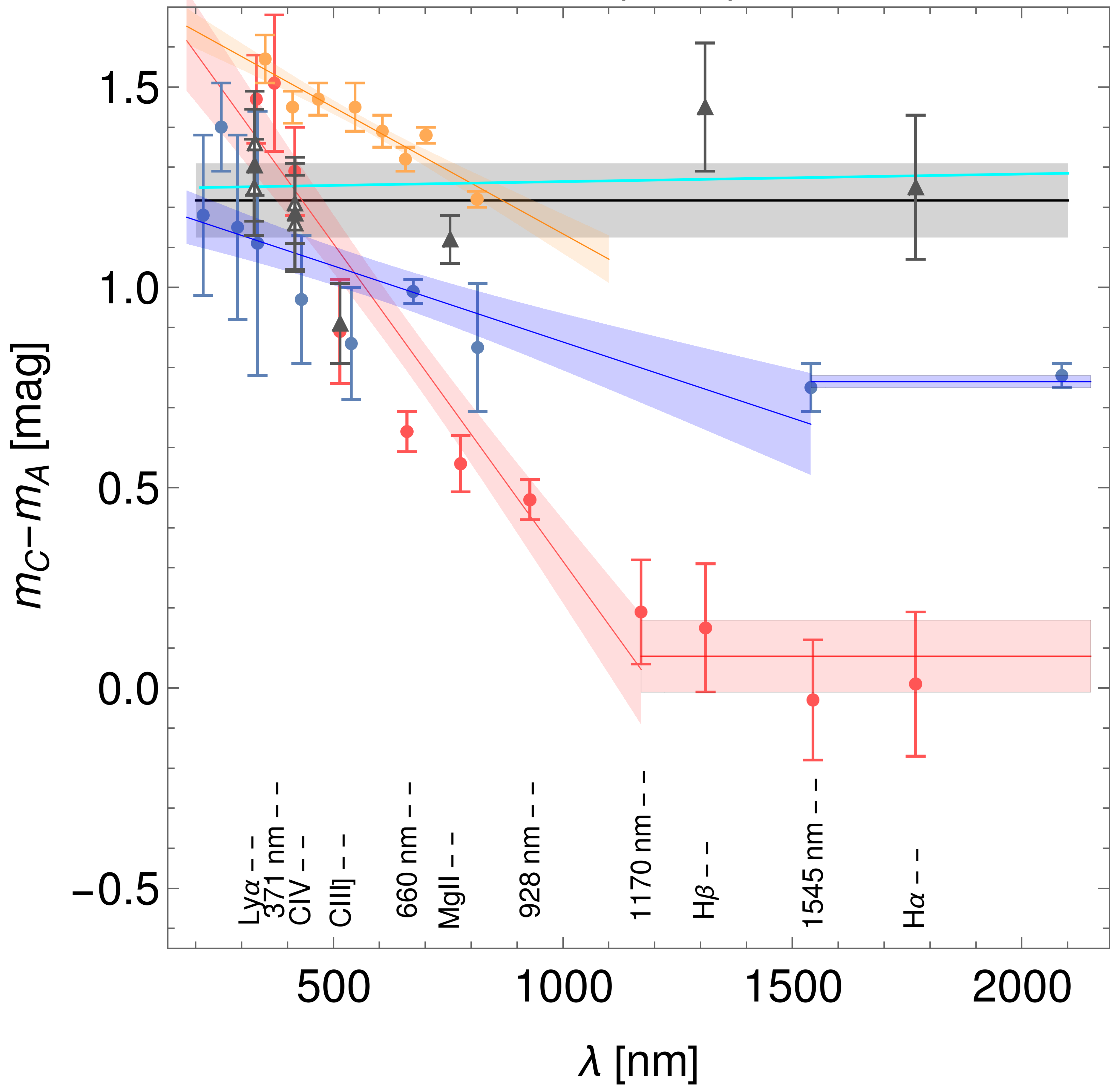}
    \includegraphics[width=0.322\textwidth]
    {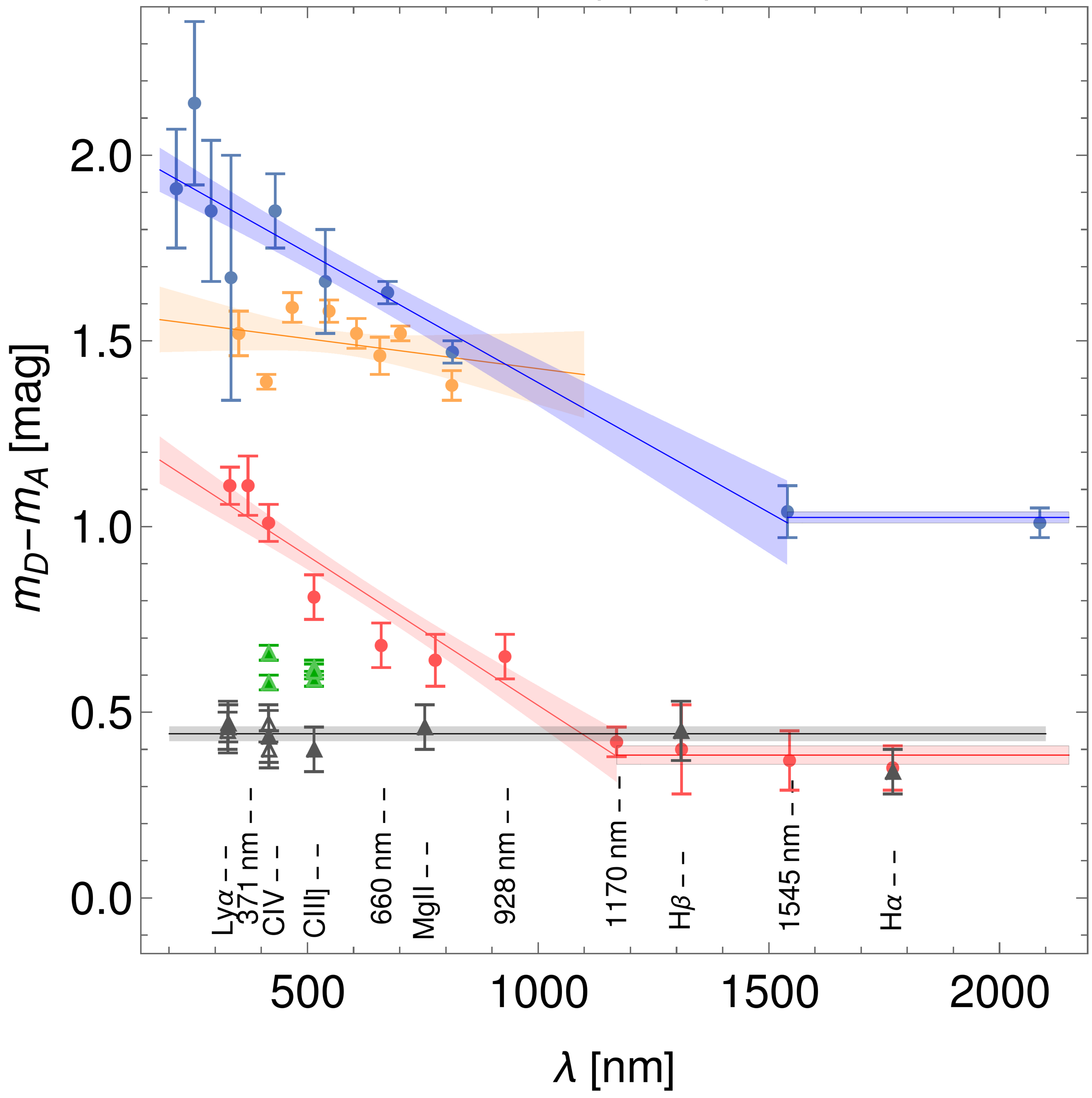}
 \caption{Magnitude differences, $\Delta m~(B-A)$ (left), $\Delta m~(C-A)$ (middle), and $\Delta m~(D-A)$ (right), as a function of wavelength for Q2237+0305. The wavelength axis is in the quasar frame. Black triangles correspond to the core emission lines without continuum. Red circles represent the continuum below the emission lines, where additional continuum measurements are included, corresponding to certain windows of 15 to 50 nm width depending on the continuum region selected (see Table \ref{table:dif_Q2237}), all
 marked with their central wavelengths ($\lambdaup_{c}$). Orange and blue circles are the continuum from \cite{mosquera2009} and CASTLES database, respectively. Green triangles in the $\Delta m~(D-A)$ plot correspond to magnitude differences obtained by \cite{sluse2011} for the C{\footnotesize{III}}] and C{\footnotesize{IV}} emission lines and the solid cyan line in the $\Delta m~(C-A)$ plot shows our calculated extinction curve (see text). }
    \label{fig:BA_md}
\end{figure*}

By analyzing this sample of emission lines for all the quasar images, we confirmed microlensing effect and chromaticity between image pairs for this system with respect to image A, in agreement with previous studies \citep{2011Mosquerab,sluse2011,mediavilla2015,munoz2016,vives-arias2016,braibant2016,2021hutse,savic2024}. The magnitude differences between quasar image pairs for the continuum and core of the selected emission lines ($\Delta m^{cont}$ and $\Delta m^{core}$) can be seen in Figure \ref{fig:BA_md} as red and gray data, respectively. In the case of Ly$\alpha$ and C{\footnotesize{IV}} emission lines, we detected prominent absorption lines in 
their cores; thus, for these two lines, the magnitudes, $\Delta m^{core}$, were obtained using the $\Delta m$ mean values of those two small integration windows. 
We applied the median estimation for all the $\Delta m^{core}$, establishing this as our no-microlensing baseline, where $\Delta m^{core}_{BA}=0.47 \pm 0.03$ mag; $\Delta m^{core}_{CA}=1.22\pm 0.09$ mag and $\Delta m^{core}_{DA}=0.44\pm0.02$ mag.

In addition, we used the multi-epoch photometry data described in Section 2.2 corresponding to the differential magnitude measurements of the continuum, $\Delta m^{cont,c}$ (from the CASTLES database) and $\Delta m^{cont,m}$ (from \citealt{mosquera2009}), along with our \texttt{X-shooter} continuum, $\Delta m^{cont}$, to subtract them from the median measurements of the line cores $\Delta m^{core}$ measured from \texttt{X-shooter} data (see Table \ref{table:chromatic_microlensing}).
We found that the chromatic microlensing signal is different for the three epochs, which allows us to obtain three independent measurements of chromaticity in Q2237+0305 for each image pair.
In the case of $B-A$ and $D-A$, the chromatic effect behaves as usual, presenting high magnification in the bluest part and low or null magnification in the reddest part for every continuum measurement, with a limit in $\lambdaup \sim 1170$ nm in which chromaticity is no longer present. However, in the case of $C-A$ we have the opposite situation when comparing our no-microlensing baseline with $\Delta m^{cont,c}$, $\Delta m^{cont,m}$, and $\Delta m^{cont}$ (notice in Figure \ref{fig:BA_md} how the red and blue continuum lines in the NIR part of the spectrum are clearly separated from the line cores at different epochs). This result could be explained with a combination of chromatic microlensing present in the optical continuum in image $A$ plus an additional microlensing magnification of $C$ of around -1.2 mag observed in the NIR continuum. Years ago, the same scenario was proposed by \cite{motta2017} for the HE1104-1805 gravitational lens system, between $A$ and $B$ quasar images. In this scenario,  both $A$ and $C$ quasar images would be affected by a microlensing phenomenon with different magnifications.

\begin{table}[!t]
\centering
\caption{Chromatic microlensing in Q2237+0305.}
\label{table:chromatic_microlensing}
\small
\renewcommand{\arraystretch}{1.3}
\begin{tabular}{llll} \hline\hline
Images 
& \makecell{$\Delta m^{cont,c}$  $-\ \Delta m^{core}$}
& \makecell{$\Delta m^{cont,m}$  $-\ \Delta m^{core}$}
& \makecell{$\Delta m^{cont,x}$  $-\ \Delta m^{core}$} \\

& (mag) & (mag) & (mag) \\ \hline\hline

\multirow{3}{*}{$B-A$} 
& $1.07 \pm 0.06$ & $1.13 \pm 0.08$ & $0.23 \pm 0.06$ \\ 
& $0.93 \pm 0.05$ & $1.08 \pm 0.05$ & $0.12 \pm 0.04$\\ 
& $0.75 \pm 0.05$ & $1.03 \pm 0.07$ & $-0.02 \pm 0.05$\\ \hline

\multirow{3}{*}{$C-A$} 
& $-0.10 \pm 0.11$ & $0.34 \pm 0.10$ & $0.17 \pm 0.13$\\ 
& $-0.17 \pm 0.10$ & $0.22 \pm 0.09$ & $-0.13 \pm 0.12$\\ 
& $-0.26 \pm 0.11$ & $0.07 \pm 0.10$ & $-0.51 \pm 0.12$\\ \hline

\multirow{3}{*}{$D-A$} 
& $1.42 \pm 0.05$ & $1.09 \pm 0.06$ & $0.62 \pm 0.05$\\ 
& $1.28 \pm 0.05$ & $1.06 \pm 0.04$ & $0.47 \pm 0.04$\\ 
& $1.12 \pm 0.05$ & $1.02 \pm 0.05$ & $0.27 \pm 0.04$\\ \hline\hline
\end{tabular}

\vspace{0.2cm}
\begin{flushleft}
\footnotesize{\textbf{Notes.}}
Each row of every cell corresponds to a measurement at the corresponding $\lambda=327,~516,$ and $754$ nm. $\Delta m^{cont,c}$, $\Delta m^{cont,m}$, and $\Delta m^{cont,x}$ are the magnitude differences for the continuum of CASTLES, \cite{mosquera2009}, and \texttt{X-shooter}, respectively.
\end{flushleft}
\end{table}

\subsection{Estimation of dust extinction by the lens galaxy}

We calculated the extinction produced by dust in the lens galaxy on the $C$ image following the procedure detailed in Section 3.3 and we found that the values obtained for $\Delta{m}^{\text{core}}$ contradict previous works \citep{falco1999,eigenbrod2008}, which reported differential extinction affecting the quasar images. Based on broadband measurements, \cite{falco1999} found $\Delta E_{A}(B-V)=0.07\pm0.03$ mag, $\Delta E_{B}(B-V)=0$ mag, $\Delta E_{C}(B-V)=0.18\pm0.03$ mag, and $\Delta E_{D}(B-V)=0.17\pm0.03$ mag and a total $E(B-V)$ of $0.13\pm0.04$ mag.
Additionally, \cite{eigenbrod2008} reported a reddening in image $C$ relative to images $A$ and $B$ when estimating the parameters of differential extinction, where $A_{V}(C)-A_{V}(A)\simeq A_{V}(C)-A_{V}(B)$, falling within the range of $0.1-0.3$ mag. 
As was mentioned in Section 3.3., we used the extinction law of \cite{cardelli1989} assuming the canonical value $R_V=3.1$ for the Milky Way with a variable extinction ($\Delta E$) and magnification ratio ($\mu_1/\mu_2$).
Using the magnitude differences measured in the core of the emission lines, we find that the extinction difference for the $C-A$ image pair in our analysis is $\Delta E(B-V)_{CA}=0.04$ mag, with the corresponding extinction curve shown in Figure \ref{fig:BA_md} as a solid cyan line. Taking into account its negligible slope, we confirm that no chromatic variations are observed in $\Delta{m}^{\text{core}}$ for the $C-A$ image pair associated with extinction, for either pair $B-A$ or $D-A$. 
As we have extended the sample of emission lines toward the blue (red) with the Ly$\alpha$ (H$\beta$ and H$\alpha$) lines,  we could broaden the sample regarding the $C-A$ extinction measurements obtained by \cite{eigenbrod2008}. This fact, added to a potential increase in the spectral resolution, may explain the difference between the two results. Thus, we conclude that the presence of extinction in the images $A$ and $C$ of the quasar Q2237+0305 can be safely ruled out.

\subsection{Constraints on the accretion disk parameters of Q2237+0305}

\begin{figure}[]
    \hspace*{-0.25cm}
    \includegraphics[width=0.5\textwidth, height=0.4\textwidth]{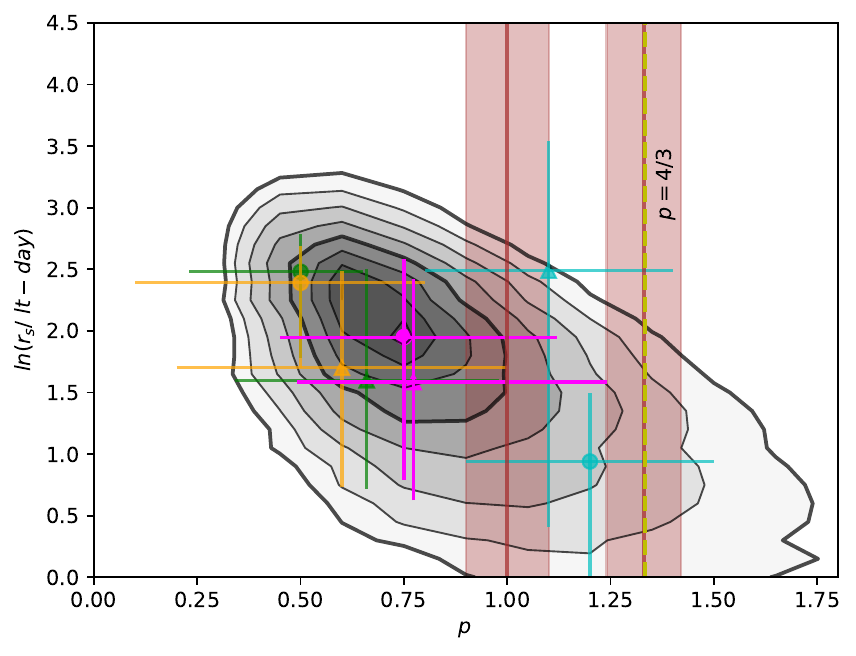}
    \caption{2D-PDF from the joint likelihood estimation using chromatic microlensing measurements from Table \ref{table:chromatic_microlensing} for Q2237+0305. Black contours represent levels from 0.25$\sigma$ to 2$\sigma$, with the corresponding 1$\sigma$ and 2$\sigma$ levels marked in bold. The magenta triangle (circle) corresponds to our expected (maximum) values, while green, orange, and blue cyan triangles (circles) show previous estimations from \cite{munoz2016}, \cite{jimenez-vicente2014}, and \cite{eigenbrod2008iii}, which correspond to the expected (maximum) estimations in the case of the first two authors, while the \cite{eigenbrod2008iii} results denote their Bayesian estimates without (with) velocity priors. The yellow line represents the canonical $p=4/3$ thin-disk model and the solid and dashed red ones (almost coincident with the yellow line) are results from \cite{Goicoechea2020} from their single and joint analyses, respectively (see text).}
    \label{fig:contour_combined}    
\end{figure}

For the estimation of the accretion disk size and its temperature profile, we compare our simulations detailed in Section 4 with the results of Table \ref{table:chromatic_microlensing}, corresponding to the microlensing magnitude differences between the continuum and emission line at wavelengths centered on the Ly$\alpha$, C{\footnotesize{III]}} and Mg{\footnotesize{II}} lines for all image pairs. The results of the joint 2D-PDF for the nine different measurements can be seen in Figure \ref{fig:contour_combined}. Our corresponding maximum and estimated values are marked as a magenta circle and triangle, respectively, combined with other previous estimates from \cite{jimenez-vicente2014}, \cite{munoz2016} and \cite{eigenbrod2008iii} presented as orange, green, and cyan points and \cite{Goicoechea2020} results in solid and dashed red lines. We obtained $\ln{r_{s}}=1.95^{+0.63}_{-1.16}$ and 
$\ln{r_{s}}=1.59^{+0.83}_{-0.96}$ (lt-day) for the maximum and expected values of the radius, which scale with the microlens mass as $r_{s}=7.0^{+1.9}_{-3.2}$ $\sqrt{M/0.3M_{\odot}}$ and $r_{s}=4.9^{+2.3}_{-2.6}$ $\sqrt{M/0.3M_{\odot}}$ (lt-day) measured at $\lambdaup_{ref}=1026$ $\AA$. The corresponding values of the power-law index are $p=0.75^{+0.37}_{-0.30}$ and $p=0.77^{+0.47}_{-0.28}$, respectively. Notice that our results of ln ${r_{s}}$ and $p$ are compatible within a 1$\sigma$ level with those obtained in \cite{jimenez-vicente2014} and \cite{munoz2016} using $\lambdaup=1026$ and $1736$ $\AA$ as reference wavelengths, with an average microlens mass of $1M_{\odot}$ and $0.3M_{\odot}$, respectively. These values were scaled to our reference wavelength and mass of the microlenses for a proper comparison. Although \cite{munoz2016} also tested other non-Gaussian brightness profiles, we highlight that no significant differences arise in the ${(\ln r_{s},p)}$ probability distribution for this lens system.

Also, Figure \ref{fig:contour_combined} 
shows that the thin-disk $p=4/3$ value \citep{Shakura1973} (yellow line) and the results from \cite{Goicoechea2020} (red lines) and \cite{eigenbrod2008iii} (cyan points) are larger than our estimates, although they are still consistent within $2\sigma$ (black outer contour of our 2D-PDF). Nevertheless, \cite{eigenbrod2008iii} provide a value for the size of the accretion disk using a velocity prior (cyan circle), which is almost at a $2\sigma$ level from our estimates and a temperature profile that is shallower than ours, but within our measured uncertainties. Without this velocity prior (cyan triangle) their estimation of the size $\ln{r_{s}}=2.49^{+1.05}_{-2.08}$ (lt-day) (transformed from their $\lambdaup = 2125 \AA$ to our $\lambdaup_{ref}$ and scaled from $1$ to $0.3M_{\odot}$), provides a value in better agreement within the errors with respect to our estimation and those of \cite{jimenez-vicente2014} and \cite{munoz2016}.

In the case of \cite{Goicoechea2020}, they estimated the power-law profile based on two caustic crossing events from two different photometric periods in 2012-2013 and 2015-2016. However, these two events were treated differently following two approaches in their calculations. For the 2012-2013 event, they could only use observations from the Maidanak Telescope (MT) in the VRI filters, giving as a result $p=1.0 \pm 0.1$ (solid red line). For the 2015-2016 event, they jointly considered observations from the Liverpool Telescope (LT) and MT, from five different filters (gVrRI). Thus, in their second approach, they used these data plus 2012-2013 event data, obtaining $p=1.33 \pm 0.09$ with a final 10\% precision (dashed red line). As seen in Figure \ref{fig:contour_combined}, when the authors consider both events, the power-law index is almost equal to the standard thin-disk value, although they could not obtain any value of $\chi^{2}/d.o.f$ $\lesssim 1$ for this joint analysis. For this reason, they claim that a more detailed analysis of the two events, including simulations of the accretion disk size, may still have an impact on their estimations. In general, both of their results are more consistent with those found by \cite{eigenbrod2008iii}, but their result when using the 2012-2013 event ($p=1.0$) is also in agreement with our findings.

Finally, we obtain the half-light radius for the maximum and expected values in the Q2237+0305 system as $R_{1/2}=8.26^{+2.24}_{-3.78}$ $\sqrt{M/0.3M_{\odot}}$ (lt-day) and $R_{1/2}=5.78^{+2.71}_{-3.07}$ $\sqrt{M/0.3M_{\odot}}$ (lt-day), respectively. These values are in concordance with previous estimates from \cite{mediavilla2015, jimenez-vicente2014, munoz2016}, whose works have typically obtained values between $R_{1/2}=3-8.5$ $\sqrt{M/0.3M_{\odot}}$ (lt-day) and also from \cite{eigenbrod2008II}, whose estimations of the source size yield a value of $R_{1/2}\approx 3$ (lt-day). In addition, works from \cite{2011Blackburne} and \cite{fian2016,fian2018} for different systems also agree with the values found in this work, with an average half-light radius of $\approx3.4$, $3.7$ and $2.7$ (lt-day), respectively. All these values contrast with the results from \cite{2008morgan} for the lens system PG 1115+080, with an average value of $R_{1/2}\approx9$ (lt-day), although in this case the authors assume a thin-disk model with $p=4/3$ in their calculations. Expanding the analysis to a larger sample of lens systems and applying multiple methodological approaches would help reduce measurement uncertainties and systematic differences among studies employing different techniques.

{\subsection{BLR microlensing}}

The matched core profiles for all the emission lines are shown in Figure \ref{fig:lines_matched_Q2237} with the corresponding normalization values. In the case of Ly$\alpha$ and C{\footnotesize{IV}}, absorption lines in their cores introduce uncertainties in the normalization. As an alternative, we used the maximum value of the bluest peak, which typically holds the highest flux counts. Regarding the profiles, slight differences are observed in the red wing of Ly$\alpha$ $\sim \lambdaup[329.0:331.5]$ nm and in the Ne{\footnotesize{V}} region $\sim \lambdaup[332.0:336.0]$ nm, being brighter in $A$ and $D$, which can be attributed to either microlensing effect in the continuum emission or an unsuccessful core emission line match, due to the absorption line in its core.
In C{\footnotesize{IV}}, small differences are observed in the blue wing ($\sim\lambdaup[412.8:414.6]$ nm), with the C component more affected than others, probably related to an absorption region, and larger ones in the red wing ($\sim\lambdaup[416.2:416.7]$ nm), the latter possibly attributable to the microlensing effect \citep{fian2021}. The characteristics of the C{\footnotesize{IV}} profile found in this work are also reported by \cite{eigenbrod2008iii}, mentioning that the BEL is symmetrically magnified by the microlensing effect. To the right of C{\footnotesize{IV}} we can see that the He{\footnotesize{II}} profiles are matching very well for the different images, except for the component $C$, likely due to its low S/N.

Moreover, no significant differences are observed in C{\footnotesize{III}}], with the exception of the Al{\footnotesize{III}} line ($\sim\lambdaup[495:504]$ nm), where the $C$ component presents an increase relative to the other images. This could be a sign of microlensing in the BEL region affecting this line or due to a low S/N, which cannot be discarded. Similarly, the blue wing of Mg{\footnotesize{II}} shows a good matching, except for quasar image $C$, whose deviation could be also caused by the low S/N of the spectrum at this wavelength. The differences in the red wing of this emission line ($\sim\lambdaup[756.4:758.4]$ nm) are possibly caused by a poor telluric correction or by the presence of microlensing in the underlying pseudo-continuum of FeII \citep{Vestergaard2001}.

\begin{figure}[h!]
    \hspace*{0.25cm}
    \includegraphics[width=0.4\textwidth, height=0.4\textwidth]{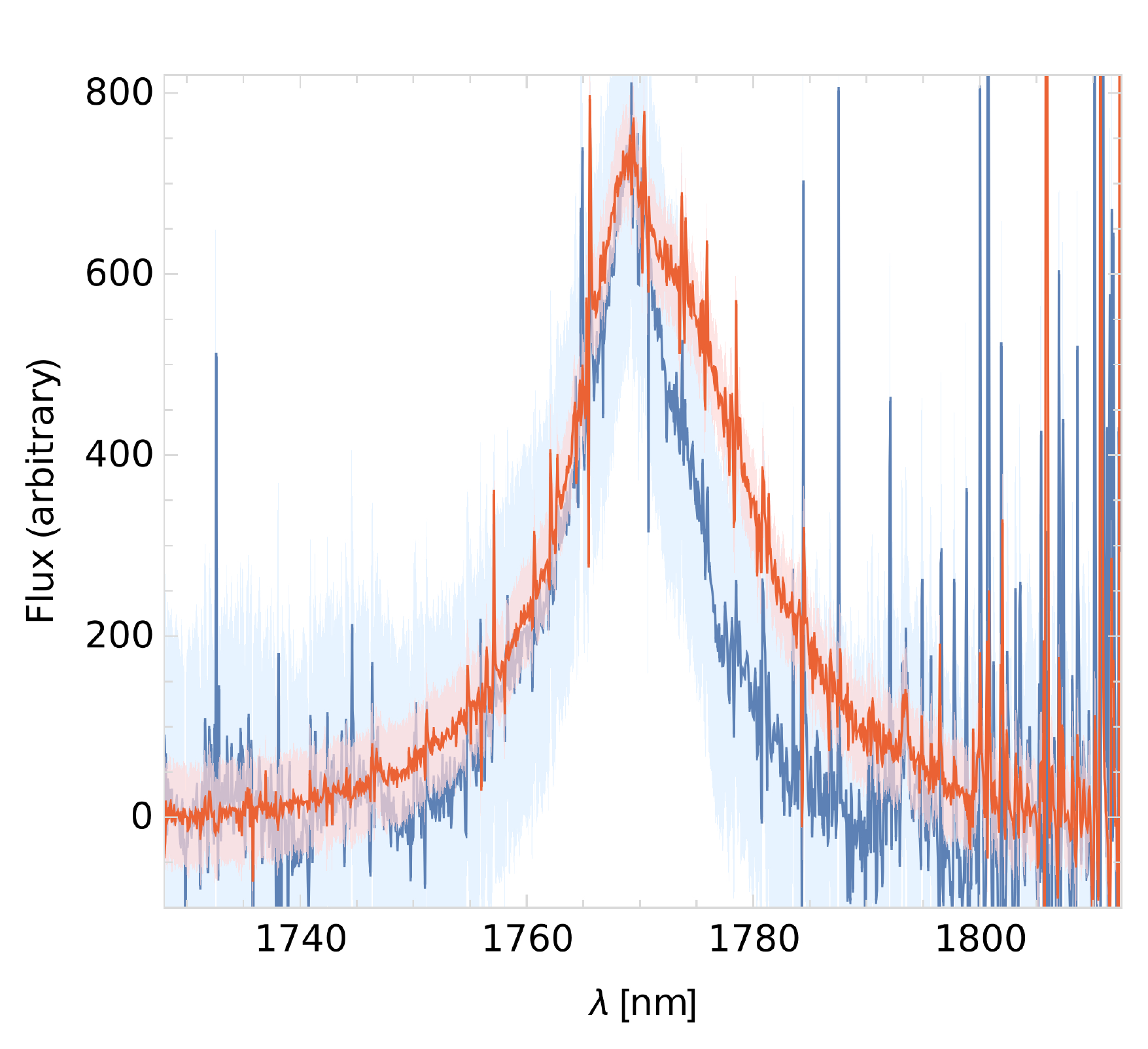}
    \caption{Comparison of the H$\alpha$ line profile between images $A$ (red line) and $C$ (blue line). Their respective $\pm1 \sigma$ total flux errors are also shown as red and blue bands, respectively, for each image.}
    \label{fig:halpha_AC}    
\end{figure}

For the H$\beta$ and H$\alpha$ line profiles, we see that those of $C$ are narrower than the other images in the regions $\sim\lambdaup[1314.5:1322.0]$ nm and $\sim\lambdaup[1773.0:1783.0 ]$, respectively. To quantify how narrow the lines are, we fit broad and narrow Gaussian components to those emission lines, estimating their FWHM ratio, i.e., FWHM$_{H\beta}$: $A/C = 2.55$; $B/C = 2.26$; $D/C = 2.24$, and FWHM$_{H\alpha}$: $A/C$ = 1.84; $B/C = 1.34$; $D/C = 1.19$. These differences could also be due to a low S/N affecting the BEL (see Figure \ref{fig:halpha_AC} and the comparison of the H$\alpha$ line between images A and C, with the total error calculated as described in Section 3.1). However, if this is solely attributed to a poor signal, the H$\alpha$ line, which usually has an improved S/N compared to H$\beta$, should not exhibit the same behavior. Instead, both lines appear to be affected in a similar way, and this feature is observed only in the red wings of the lines. In this context, a plausible alternative explanation is the presence of a persistent microlensing effect in the BLR. Note that other authors have also found the same behavior for other lens systems. Specifically, according to \cite{melo2021}, they observed a stretching of the red wing in the H$\alpha$ line of image $A$ for the lensed system WGD2038-4008. These spectra were also taken with the \texttt{X-shooter} instrument. As a second example, \cite{braibant2014} presented a similar behavior for the blue wing in the H$\alpha$ line of the component $D$ for the system HE0435-1223, although they attributed this effect to a Keplerian disk originating in H$\alpha$. Afterward, \cite{2019hutse} provided simulations of the H$\alpha$ line profile, which favored a Keplerian disk with winds for the BLR of this system.

In addition, it is worth mentioning that the Q2237+0305 data obtained by the CASTLES survey \citep{falco1999} were taken almost 20 years before the \texttt{X-shooter} spectra. At that time, the $C$ component was the third brightest image of the system. However, due to the recurrent phenomena associated with microlensing effects, in our \texttt{X-shooter} epoch the $C$ component is the faintest. 
We conclude, then, that to either confirm or reject the presence of microlensing in the BLR for this range of the spectrum we would need more accurate spectra to reach a higher S/N in this band and to compare the different epochs.

Finally, for the emission lines detailed above, we measured the magnitude differences in the wings of the BELs relative to the core of each individual emission line, with $\Delta{m}^{red}=\Delta{m}^{red~wing}-\Delta {m}^{core}$ and $\Delta{m}^{blue}=\Delta{m}^{blue~wing}-\Delta {m}^{core}$ for each image pair following the procedure described by \cite{guerras2013}. To consider our estimated values as significant detections, we established a threshold of $1.5\cdot\sqrt{(\sigma_{core}^2 + \sigma_{wing}^2})$ for the red and blue wings, independently. The differential magnitude measurements for the wings are presented in Table \ref{table:dif_Q2237} (denoted as blue and red wings) and the main calculated features are listed as follows:

\begin{enumerate}
    \item $B-A$: In {Ly$\alpha$} and CIV we found mainly marginal evidence of microlensing, with $\Delta{m}^{blue}_{Ly\alpha}=0.09\pm0.1$ mag, $\Delta{m}^{red}_{Ly\alpha}=0.2\pm 0.1$ mag, respectively, and only the red wing of C{\footnotesize{IV}} with $\Delta{m}^{red}_{CIV}=0.12\pm0.1$ mag. In the Mg{\footnotesize{II}} emission line, we found the red wing magnified, with $\Delta{m}^{red}_{MgII}=0.17\pm0.08$ mag, whereas this value could be biased due to the iron pseudo-continuum (see text). The rest of the lines do not show significant differences considering the associated uncertainties.

    \item $C-A$: Differences are mainly observed in the red wing. For Ly$\alpha$ and C{\footnotesize{IV}} the values are $\Delta{m}^{red}_{Ly\alpha}=0.33\pm0.2$ mag and $\Delta{m}^{red}_{CIV}=0.46\pm0.19$ mag. Also, high values are present in both $\Delta{m}^{red}_{H\beta}=1.20\pm0.23$ mag and $\Delta{m}^{red}_{H\alpha}=0.66\pm0.26$ mag, respectively (see Figure \ref{fig:lines_matched_Q2237} and the corresponding discussion in the text above). In the case of Mg{\footnotesize{II}}, both wings are magnified equally: $\Delta{m}^{blue}_{MgII}=$ $\Delta{m}^{red}_{MgII}=0.36\pm0.08$ mag.

    \item $D-A$: Similar features appear to those for $C-A$, but with smaller differences in their red wings. In Ly$\alpha$ we found marginal evidence, $\Delta{m}^{red}_{Ly\alpha}=0.09\pm0.07$ mag, while in C{\footnotesize{IV}}, $\Delta{m}^{red}_{CIV}=0.12\pm0.07$ mag, and in Mg{\footnotesize{II}}, $\Delta{m}^{red}_{MgII}=0.24\pm0.08$ mag. \cite{sluse2011} report differences in the BLR of C{\footnotesize{IV}} and C{\footnotesize{III}}], obtained for three different observation periods. It can be seen that their result of $\Delta m^{blue}$ for C{\footnotesize{IV}} varies between $0.25\pm0.03$ mag and $0.08\pm0.03$ mag and $\Delta m^{red}$ between $0.61\pm0.03$ mag and $0.52\pm0.03$ mag (from P1 to P3), which is consistent with what is found in this work within the uncertainties. Finally, the broad component of H$\alpha$ in the red wing also shows a value of $\Delta{m}^{red}_{H\alpha}=0.21\pm0.07$ mag.

\end{enumerate}

\section{Conclusions}
In this work, we used the \texttt{X-shooter} data of the quasar lensed system Q2237+0305, which covers a wider wavelength range compared to other available data in the literature, and inspected new observations of the Ly$\alpha$, H$\alpha$, and H$\beta$ emission lines. With these new spectra and literature data from other epochs, we could estimate the microlensing differential magnitude measurements along with the extinction values. We also inferred a sample of chromatic microlensing measurements to provide an improved estimation of the accretion disk parameters. As a summary, the main results are outlined as follows:

\begin{enumerate}

    \item We analyzed the differential extinction 
    from the three image pairs: $B-A$, $C-A$, and $D-A$, including the NIR emission lines in our analysis, which helped us to extend the coverage for a more accurate estimation. No significant deviations were found from the median of the lines defining the no-microlensing baseline in images $B$ and $D$. We specifically applied an extinction curve fit to image $C$ due to its higher dispersion in the median of the lines, yielding an extinction coefficient of $\Delta E = 0.04$ mag. Our result suggests minimal extinction for image $C$, consistent with a negligible dust contribution. This differs from previous reports that used shorter wavelength ranges and suggested positive extinction levels (e.g., \citealt{falco1999, eigenbrod2008}).
    
    \item We obtained the magnitude differences between the emission-line cores and their underlying continuum for different epochs of observations. As these magnitude differences increase toward bluer wavelengths, we confirmed the presence of chromatic microlensing in images $B$, $C$, and $D$ relative to $A$. Moreover, we can conclude that this chromatic behavior disappears beyond $\lambdaup =1170$ nm in all cases, although the $C-A$ case remains less conclusive due to the possible presence of simultaneous microlensing in both images $A$ and $C$, which produces higher values between the continuum and the baseline defined by the line cores.
    
    \item From a Bayesian study based on the comparison of the microlens signal ($\Delta
    m$) -- estimated from \texttt{X-shooter} data and previous studies -- with the probabilistic predictions of microlensing magnification maps, we estimated the accretion disk parameters (size, $r_s$, and its power-law index, $p$) ruled by the observed chromatic microlensing effect between the three considered image pairs.
    With this statistical approach, we determined both of the maximum likelihood estimates (expected values): $p=0.75^{+0.37}_{-0.30}$ ($0.77^{+0.47}_{-0.28}$) and $r_{s} (\lambdaup_{ref}=1026~\AA)=7.0^{+1.9}_{-3.2}$ ($4.9^{+2.3}_{-2.6}$) $\sqrt{M/0.3M_{\odot}}$ (lt-day). Our estimations are in good agreement with previous results from \cite{jimenez-vicente2014} and \cite{munoz2016}, which used data from six epochs at five different wavelengths (narrow-band filters) and were based on the same analysis method. Thus, our work adds an additional constraint on the accretion disk parameters of the system Q2237+0305, confirming a deviation of this system from the standard accretion disk model with $p=4/3$, although presenting source size values consistent with other previous estimates.

    \item Finally, we calculated the microlensing presence in the BELs comparing the differential magnitude in the cores with the differential magnitude in the wings for every emission line. As a result, magnification differences mainly arise in Ly$\alpha$ and C{\footnotesize{IV}}, being significant only in the red wings for the pairs $C-A$ and $D-A$. In the case of the Mg{\footnotesize{II}} line, a confirmed magnification in the BELs is subject to major biases given the low S/N and the possible effect of the iron pseudo-continuum (also discussed in \citealt{fian2021}). There is also a significant difference present in the $C-A$ pair for the red wing of H$\alpha$ and H$\beta$ lines compared to the other images. While beyond the scope of this work, comparing these results with analogous systems remains a key objective for future studies.
\end{enumerate}

\begin{acknowledgements}
We thank the anonymous referee for the useful comments and suggestions. R.J.N. and V.M. acknowledge support from ANID FONDECYT Regular grant number 1231418 and Centro de Astrof\'{\i}sica de Valpara\'{\i}so CIDI 21. A.E.G. acknowledges support from project ANID Fondecyt Postdoctorado with grant number 3230554. E.M. acknowledges support from grants PID2020-118687GB-C31 and PID2020-118687GB-C33, financed by the Spanish Ministerio de Ciencia e Innovación/Agencia Estatal de Investigación MCIN/AEI/10.13039/501100011033.
\end{acknowledgements}

\bibliographystyle{aa} 
\bibliography{biblio.bib} 
\clearpage
\appendix
\onecolumn
\section{Long tables}
\begin{table*}[!h]
\centering
\caption{Estimated magnitude differences for Q2237+0305 from the continuum and core of \texttt{X-shooter} spectra.}
\renewcommand{\arraystretch}{1.4}
\begin{tabular}{lccccc}

\hline Region & $\lambdaup_{c}$ & Integration window $^{\mathrm{}}$  & $m_{B}-m_{A}{ }^{\mathrm{}}$  & $m_{C}-m_{A}{ }^{\mathrm{}}$  & $m_{D}-m_{A}{ }^{\mathrm{}}$ \\
&  $(\mathrm{nm})$& $(\mathrm{nm})$ & $(\mathrm{mag})$ & $(\mathrm{mag})$ & $(\mathrm{mag})$\\

\hline\hline  Continuum& 331.9 & $317.3-320.5/342.5-346.5 $   & $0.74 \pm 0.06$ & $1.47 \pm 0.11$  & $1.11 \pm 0.05$ \\ 
&                  371.0 & $343.0-349.0 / 394.0-399.0$  & $0.73 \pm 0.10$ & $1.51 \pm 0.17$ & $1.11 \pm 0.08$\\
&                 416.0 & $394.8-398.9 / 456.8-462.2$   & $0.64 \pm 0.06$ & $1.29 \pm 0.11$ & $1.01 \pm 0.05$\\
&                 514.0 & $483.7-489.1/533.6-544.4$     & $0.48 \pm 0.07$ & $0.89 \pm 0.13$ & $0.81 \pm 0.06$\\
&                 660.0 & $645.0-675.0$                 & $0.43 \pm 0.05$ & $0.64 \pm 0.05$ & $0.68 \pm 0.06$\\
&                 777.0 & $738.0-742.0 / 772.0-780.0$   & $0.42 \pm 0.06$ & $0.56 \pm 0.07$ & $0.64 \pm 0.07$\\
&                 928.0 & $860.0-890.0/985.0-1005.0$    & $0.42 \pm 0.04$ & $0.47 \pm 0.05$ & $0.65 \pm 0.06$\\
&                1170.1 & $1085.0-1095.0/1270.0-1286.0$ & $0.31 \pm 0.11$ & $0.19 \pm 0.13$ & $0.42 \pm 0.04$\\
&                1310.9 & $1278.8-1294.4/1340.0-1343.0$ & $0.33 \pm 0.17$ & $0.15 \pm 0.17$ & $0.40 \pm 0.12$\\
&                1545.0 & $1450.0-1640.0$               & $0.38 \pm 0.15$ & $-0.03 \pm 0.15$ & $0.37 \pm 0.08$\\
&                1768.6 & $ 1725.0 - 1733.0/1807.5 -1810.0$ & $0.39 \pm 0.11$ & $0.01\pm 0.18$ & $0.35 \pm 0.06$\\

\hline\hline Lines & Ly$\alpha~\lambdaup121.6$    & $326.5-327.2$ & $0.46 \pm 0.06$ & $1.25 \pm 0.12$ & $0.45 \pm 0.05$\\
&                                             & $328.0-328.5$ & $0.60 \pm 0.06$ & $1.36 \pm 0.13$ & $0.47 \pm 0.05$\\
&                 Ly$\alpha $ blue~wing       & $324.1-326.0$ & $0.62 \pm 0.06$ & $1.42 \pm 0.10$ & $0.51 \pm 0.05$\\
&                 Ly$\alpha$  red~wing        & $329.1-331.3$ & $0.73 \pm 0.06$ & $1.64 \pm 0.10$ & $0.55 \pm 0.05$\\

&                C IV $\lambda 154.9$       & $415.0-416.0$ & $0.52 \pm 0.06$ & $1.21 \pm 0.11$ & $0.47 \pm 0.05$\\
&                                             & $416.2-416.7$ & $0.45 \pm 0.06$ & $1.16\pm 0.12$ & $0.40 \pm 0.05$ \\
&               C IV blue wing       & $412.8-414.6$ & $0.61 \pm 0.06$ & $1.39 \pm 0.10$ & $0.59 \pm 0.05$\\
&               C IV red wing        & $418.8-422.0$ & $0.70 \pm 0.06$ & $1.65 \pm 0.10$ & $0.57 \pm 0.05$\\

&               C III $\lambdaup 190.9$       & $513.0-515.5$ & $0.37 \pm 0.07$ & $0.91 \pm 0.10$ & $0.40 \pm 0.06$\\
&               C III blue wing      & $507.5-512.2$ & $0.36 \pm 0.07$ & $0.80 \pm 0.10$ & $0.42 \pm 0.06$\\
&               C III red wing      & $516.5-520.4$ & $0.45 \pm 0.07$ & $1.17 \pm 0.10$ & $0.48 \pm 0.06$\\

&               Mg II $\lambdaup 279.9$       & $753.0-756.0$ & $0.45 \pm 0.05$ & $1.12 \pm 0.06$ & $0.46 \pm 0.06$\\
&               Mg II blue wing      & $749.0-752.5$ & $0.50 \pm 0.05$ & $1.48 \pm 0.06$ & $0.45 \pm 0.06$\\
&               Mg II red wing       & $756.4-758.4$ & $0.63 \pm 0.05$ & $1.48 \pm 0.06$ & $0.70 \pm 0.06$\\

&               H$\beta~\lambdaup 486.1$    & $1307.0-1312.3$ & $0.44 \pm 0.15$ & $1.45\pm 0.16$ & $0.45 \pm 0.08$ \\
&               H$\beta$ blue wing & $1302.5-1305.8$ & $0.51 \pm 0.16$ & $1.28 \pm 0.17$ & $0.54 \pm 0.08$\\
&               H$\beta$ red wing  & $1314.5-1322.0$ & $0.52 \pm 0.16$ & $2.62 \pm 0.17$ & $0.62 \pm 0.08$\\

&                 H$\alpha~\lambdaup 656.2$ & $1766.3-1771.0$ & $0.50 \pm 0.10$ & $1.25 \pm 0.18$ & $0.34 \pm 0.06$\\
&              H$\alpha$ blue wing & $1755.0-1763.5$ & $0.46 \pm 0.09$ & $1.36 \pm 0.18$ & $0.35 \pm 0.05$\\
&              H$\alpha$ red wing  & $1773.0-1783.0$ & $0.61 \pm 0.09$ & $1.91 \pm 0.19$ & $0.55 \pm 0.05$\\ \hline
\end{tabular}
\label{table:dif_Q2237}

\end{table*}

\end{document}